# Co-existence of synchronization and anti-synchronization in Generalized Lorenz System with application to secure communications.

V. RamiyaGowse[1] ● B. Palanivel[1] ●S. Sivaprakasam[2]

**Abstract:** Synchronization and anti- Synchronization between two Generalized Lorenz Systems (GLS) coupled in a master - slave configuration is investigated. Coupling between the master and slave is enabled through a non-linear control mechanism. Synchronization between the state variables of master and slave is achieved by appropriate choice of parameters. When one of the control parameter is varied, a continuous change-over from synchronization to anti-synchronization is observed. Anti-synchronization between master and slave is achieved for two of the state variables while the third state variable exhibits synchronization. A study of encoding –decoding of messages in this system is carried out. Messages of distinct frequencies are encoded at each of the state variables of master. These messages are decoded and recovered at slave for both states of synchronization viz.: synchronized and anti-synchronized states.

___________________________

✉ **S. Sivaprakasam**
   siva@iitk.ac.in

[1] Department of Physics, Pondicherry Engineering College, Pondicherry-605 014, India
[2] Department of Physics, Pondicherry University, Pondicherry-605 014, India

**Introduction**

Significant research activities on synchronization of chaotic systems have been carried out in the recent decades. This is because of its direct application in both fundamental and applied research. A specific example of its utility is secure communications. This field of research gained attention after the demonstration of chaos synchronization by Pecora and Carroll in 1990 which had electronic circuits as the study system [1]. The possibility of controlling chaos [2] and the proposal of using such controlled chaos towards communicating information have been shown earlier [3]. Several works which includes lasers, for synchronization and its application in secure communication aspects had been reported [4-10]. Few examples of other fields of research that has a basis on chaos synchronization are neuroscience, hydrodynamics, chemical and biological systems [11-14].

In the past, various types of synchronization have been identified and the major types are complete synchronization, generalized synchronization, inverse synchronization and phase synchronization [15-19]. Both experimental and theoretical research using optical chaos in semiconductor lasers demonstrated the achievement of synchronization of various types. [20-28].

Study of laser dynamics is mutually complementary with the studies of generalized oscillators such as Rossler, Lorenz and their variants. Such studies provide a platform of flexibility and are applicable to real systems such as lasers. After the reporting of inverse synchronization in laser systems, Inverse (also referred to anti-) synchronization phenomena had been reported in coupled chaotic Lorenz oscillators [29]. Antisynchronization is a phenomenon in which the state vectors of synchronized systems have the same amplitude but opposite signs compared to those of the driving system [30]. Studies of chaotic oscillators have significance in that they constitute a relevant basis for secure communications [31-34]. Recently, inverse projective synchronization between two different hyper-chaotic systems using a non-linear control mechanism was reported [35]. Also, a general procedure has been presented towards achieving synchronization, anti synchronization and amplitude death in systems such as Lorenz, Rossler and Sprott [36-39].

In this work, a master - slave system made up of two generalized Lorenz systems (GLS) is considered. A non-linear control is used to enable coupling between the master-slave systems. Continuous change-over of synchronization to anti-synchronization is found to occur between the state variables and this is achieved by controlling one of the master parameters. Synchronization is realized by plotting the time evolution of slave's state variable against the master. For synchronization, a positive slope plot emerges. In this investigation, suitable change of a control parameter is found to gradually vary the synchronization to anti-synchronization.

Three messages of distinct frequencies are simultaneously encoded to the three state variables of the master and all the three messages are recovered at one of the state variable of the slave system. This process of encoding and decoding is found to be possible for both synchronized and anti-synchronized states.

**Theoretical Analysis**

A master - slave system is considered for investigation and they belong to a class of Generalized Lorenz chaotic System (GLS).

The master system is described as follows [40]:

$$\dot{x}_1 = a(x_2 - x_1)$$
$$\dot{x}_2 = bx_1 + dx_2 - x_1 x_3 \quad (1)$$
$$\dot{x}_3 = x_1 x_2 + cx_3$$

The slave system is described by:

$$\dot{y}_1 = a(y_2 - y_1) + u_1$$
$$\dot{y}_2 = by_1 + dy_2 - y_1 y_3 + u_2 \quad (2)$$
$$\dot{y}_3 = y_1 y_2 + cy_3 + u_3$$

Where, $a = 10 + (25/29)k, b = 28 - (35/29)k, c = -(8/3) - (1/87)k, d = k - 1$

In here, $x(t)=[x_1(t),x_2(t),x_3(t)]^T$ and $y(t)=[y_1(t),y_2(t),y_3(t)]^T$ are the state variables of the master and slave GLS systems respectively, k is the system parameter; $u_1, u_2$ and $u_3$ are the control functions for achieving simultaneous synchronization and anti-synchronization between the master and the slave system.

The error is defined as,

$$E_1 = y_1 + \sigma_1 x_1$$
$$E_2 = y_2 + \sigma_2 x_2 \quad (3)$$
$$E_3 = y_3 + \sigma_3 x_3$$

Where, $\sigma=(\sigma_1, \sigma_2, \sigma_3)$ are the control parameters of the system, influencing the dynamics of coupling between the master and slave.

From equation (1)-(3) the following is the error dynamical system:

$$\dot{E}_1 = a(y_2 - y_1) + \sigma_1 a(x_2 - x_1) + u_1$$
$$\dot{E}_2 = \sigma_2 b x_1 + \sigma_2 d x_2 - \sigma_2 x_1 x_3 + b y_1 + d y_2 - y_1 y_3 + u_2 \quad (4)$$
$$\dot{E}_3 = y_1 y_2 + c y_3 + \sigma_3 x_1 x_2 + \sigma_3 c x_3 + u_3$$

where,

$$u_1 = u_{a1} - u_{b1} + u_{c1}$$
$$u_2 = u_{a2} + u_{b2} + u_{c2} \quad (5)$$
$$u_3 = -u_{a3} - u_{b3}$$

In which, $u_{a1} = (\sigma_2 - \sigma_1)a x_2$, $u_{b1} = (b + \sigma_3 x_3)E_2$, $u_{c1} = (\sigma_2)x_2 E_3$, $u_{a2} = (\sigma_1 - \sigma_2)b x_1$, $u_{b2} = (\sigma_1 \sigma_3 + \sigma_2)x_1 x_3$, $u_{c2} = E_3 E_1 - a E_1 - 2c E_2$, $u_{a3} = (\sigma_1 \sigma_2 + \sigma_3)x_1 x_2$, $u_{b3} = E_1 E_2$.

Substituting (5) into (4) leads to the following :

$$\dot{E}_1 = -a E_1 + (a - b - \sigma_3 x_3)E_2 + \sigma_2 x_2 E_3$$
$$\dot{E}_2 = (b - a + \sigma_3 x_3)E_1 + (d - 2c)E_2 + \sigma_1 x_1 E_3 \quad (6)$$
$$\dot{E}_3 = -\sigma_2 x_2 E_1 - \sigma_1 x_1 E_2 + c E_3$$

Now, let us consider the following Lyapunov function:

$$V = \frac{1}{2}\left(E_1^2 + E_2^2 + E_3^2\right) \tag{7}$$

Differentiating V, with respect to time (t):

$$\dot{V} = E_1\dot{E}_1 + E_2\dot{E}_2 + E_3\dot{E}_3$$

$$= -aE_1^2 + (a - b - \sigma_3 x_3)E_1E_2 + \sigma_2 x_2 E_1 E_3 + (b - a - \sigma_3 x_3)E_2 E_1 + (d - 2c)E_2^2 + \sigma_1 x_1 E_2 E_3$$

$$- \sigma_2 x_2 E_3 E_1 - \sigma_1 x_1 E_3 E_2 + cE_3^2$$

$$= -(E_1 \quad E_2 \quad E_3) Q (E_1 \quad E_2 \quad E_3)^T \tag{8}$$

Where, $Q = \begin{pmatrix} a & -(b - a - \sigma_3 x_3) & \sigma_2 x_2 \\ (b - a + \sigma_3 x_3) & (2c - d) & \sigma_1 x_1 \\ -\sigma_2 x_2 & -\sigma_1 x_1 & -c \end{pmatrix}$ (9)

To ensure that the error system is stabilized at the origin, the matrix Q needs to be positive definite. This is the case if and only if the following conditions hold:

(i) $a > 0$

(ii) $a^2 + b^2 - 2a(b - c) - ad > P^2$

(iii) $2MNP + aM^2 + (2c - d)N^2 + cP^2 > c(2ac - ad + 2ab - a^2 - b^2)$ (10)

Where, M, N and P are the upper bounds of the absolute values of variables $x_1$, $x_2$ and $x_3$ respectively.

The above conditions in eqn.(10) can be expressed in terms of inequality of k as follows:

(i) $10 + \left(\frac{25}{29}\right)k > 0$

(ii) $k^2 - 527.49k + 686.03 > P^2$

(iii) $-(0.0056k^3 + 1.84k^2 - (0.86M^2 + 1.01N^2 + 0.01P^2 - 109.8)k) > -2MNP + 10M^2 - 4.33N^2 + 2.67P^2$

Calculation of the parameters is carried out and the values obtained are : $k \cong 0.5$, $M \cong 21, N \cong 30 \,\&\, P \cong 21$, and these bounds of M, N, and P are confirmed using numerical simulations as shown in Figure.1 (a)-(c) and the results are discussed in the subsequent sections.

Hence the condition $\dot{V}<0$ is satisfied, which implies that the origin of error system is asymptotically stable. Using these set of parameters, simulations are carried out to realize the co-existence of synchronization and anti-synchronization. These results are presented in the next section.

**Co-existence of Synchronization and anti- Synchronization**

Equations (1) and (2) are numerically solved using Runge-Kutta algorithm keeping the initial values of the master and the slave system as $(x_1, x_2, x_3)^T =$ (0.999, 0.899, 0.799)$^T$ and $(y_1, y_2, y_3)^T=$ (1.0, 1.0, 1.0)$^T$ respectively with time step-size, h =0.05 and k =0.5. The nature of synchronization of the system is investigated by varying the value of the control parameter $\sigma$. Initially the value of ($\sigma_1, \sigma_2, \sigma_3$) is set as $\sigma=$ (0, 0, 0) and the dynamics of coupled master-slave GLS system is studied.

In figure.1 (a)-(c) the chaotic time evolution of the state variables of the master and the slave systems are shown. Figure 1(d)-(e) shows that after an initial transient the anti-synchronization errors $E_1$ and $E_2$ converge to zero. However, $E_3$ does not converge and rather remain strongly oscillatory in nature as shown in figure 1(f) as dash-dot line.

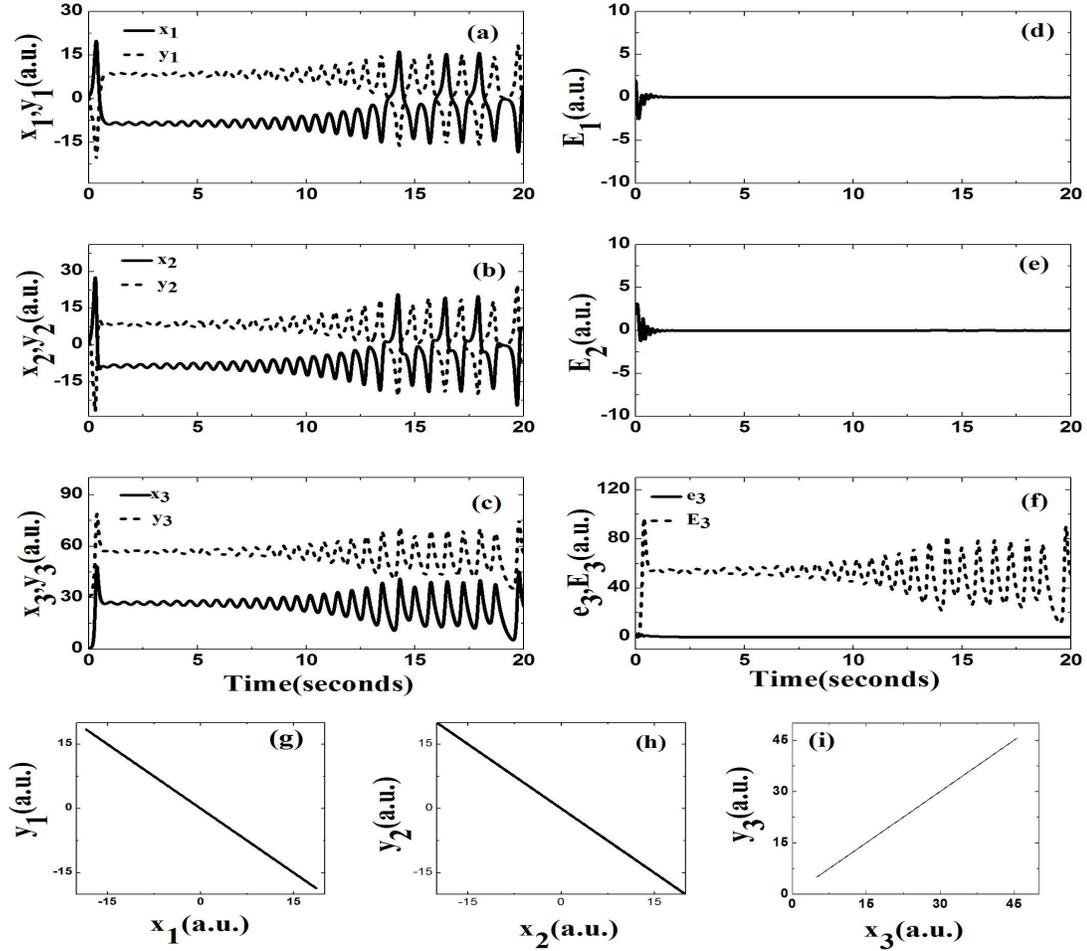

**Fig.1** (a)-(c) are the evolution of state variables $x_i$ and $y_i$, i=1,2,3 and (d)-(f) are their respective error dynamics. (g)-(i) are the synchronization plots between the state variables. The parameter σ=0.0

If the synchronization error is defined as $e_i = y_i - x_i$; where i = 3, the error converges to zero as time approaches infinity as shown in figure 1(f). This is shown as the bottom trace in solid line. Hence, it is evident that the GLS system under study simultaneously satisfies both the conditions of synchronization and anti-synchronization. The system is being stabilized by the control inputs $u_1$, $u_2$ subject to the limiting conditions $\lim_{t \to \infty} \|E_i(t)\| = 0$ and by the control input $u_3$ subject to the limiting condition $\lim_{t \to \infty} \|e_i(t)\| = 0$ ($E_i = y_i + x_i$, where $i = 1, 2$ and $e_i = y_i - x_i$; where $i = 3$) implying the co-existence of synchronization and anti-synchronization. The synchronization plots shown in figure.1 (g)-(i) indicates that the state variables, $x_1$ & $x_2$; of the master system are anti-synchronized with $y_1$ & $y_2$ of the slave system respectively whereas, the state variable, $x_3$ is synchronized with $y_3$. These simulations have been carried using the parameters obtained theoretically in the previous section. The simulations confirm the bounded values of M, N and P and thus simulations and theoretical analysis are in mutual concurrence.

The value of $\sigma$ is varied in small steps and it is observed that the state variables $x_1$ & $y_1$ and $x_2$ & $y_2$ are always anti-synchronized throughout the region of this study ($\sigma = $ 0 to1). But the nature of synchronization between $x_3$ and $y_3$ could be altered. On changing the value of $\sigma$, the slope of the synchronization plot [40] of $x_3$ vs. $y_3$ changes continuously. As the value of $\sigma$ increases, the slope of synchronization plot goes on reducing to almost zero for $\sigma = [\sigma_1, \sigma_2, \sigma_3] = [0.4, 0.4, 0.4]^T$. On further increasing the value of $\sigma$, the slope of synchronization plot becomes negative and for $\sigma = [\sigma_1, \sigma_2, \sigma_3] = [1, 1, 1]^T$, the system becomes completely anti-synchronized which is evident from the figure 2(g)-(i). In figure 2, figures 2(a) to 2(c) are the time evolution of the state variables of the master and slave. Figures 2(d) to 2(f) show their respective error dynamics. The synchronization plots are shown in figure 2(g), 2(h) and 2(i) and it is evident that all the three parameters display anti-synchronization.

In order to quantify the change -over of the sign of synchronization, the synchronization plot is plotted and the slope (m) of the fitted line and its variance ($\Delta m$) are obtained [40]. It is observed from figure 3(a)-(f) that as the value of $\sigma$ is varied from 0 to 1, slope (m) of the synchronization plot varies from +1.0 (positive

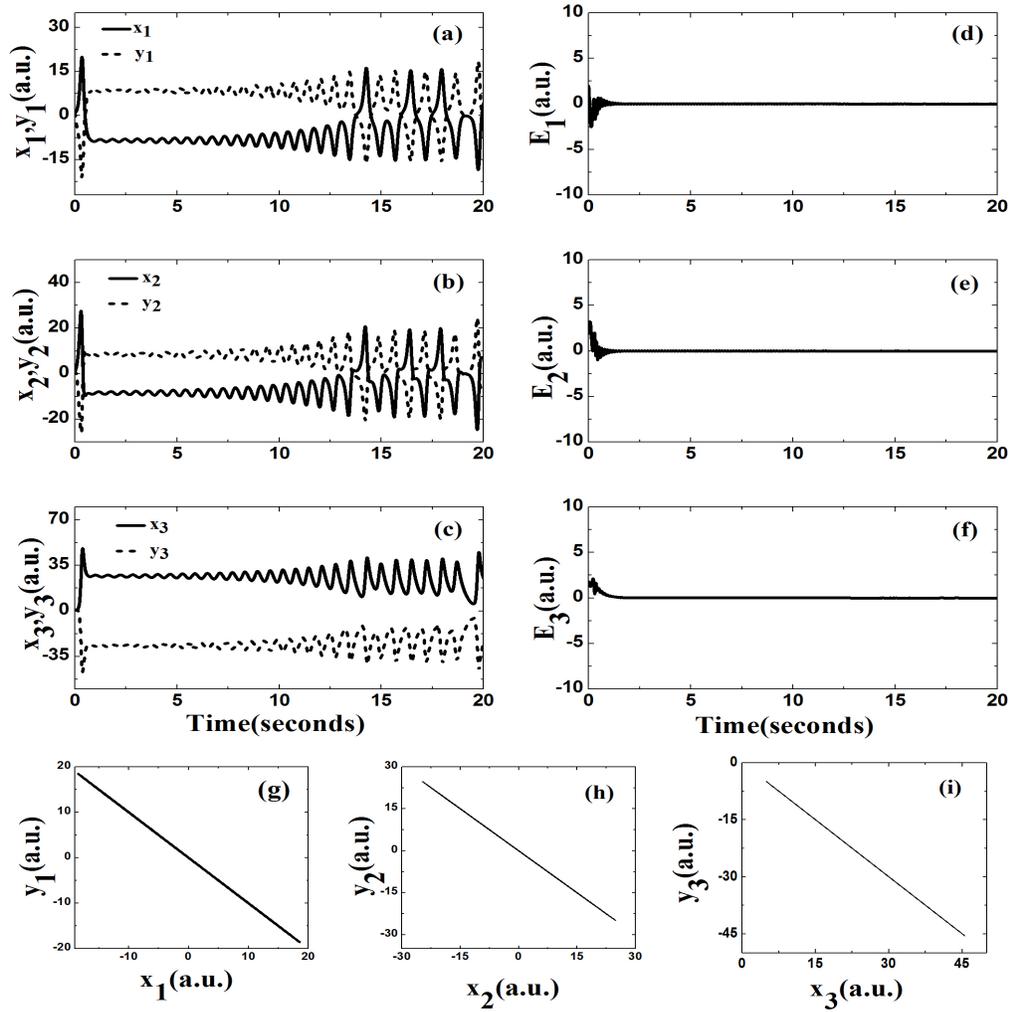

**Fig.2** (a)-(c) are the evolution of state variables $x_i$ and $y_i$, i=1,2,3 and (d)-(f) are their respective error dynamics. (g)-(i) are the synchronization plots between the state variables. The parameter $\sigma=1.0$ slope) to -1.0(negative slope), indicating that the nature of synchronization has changed from synchronization (figure3(a)) to anti-synchronization (figure(3(f)).

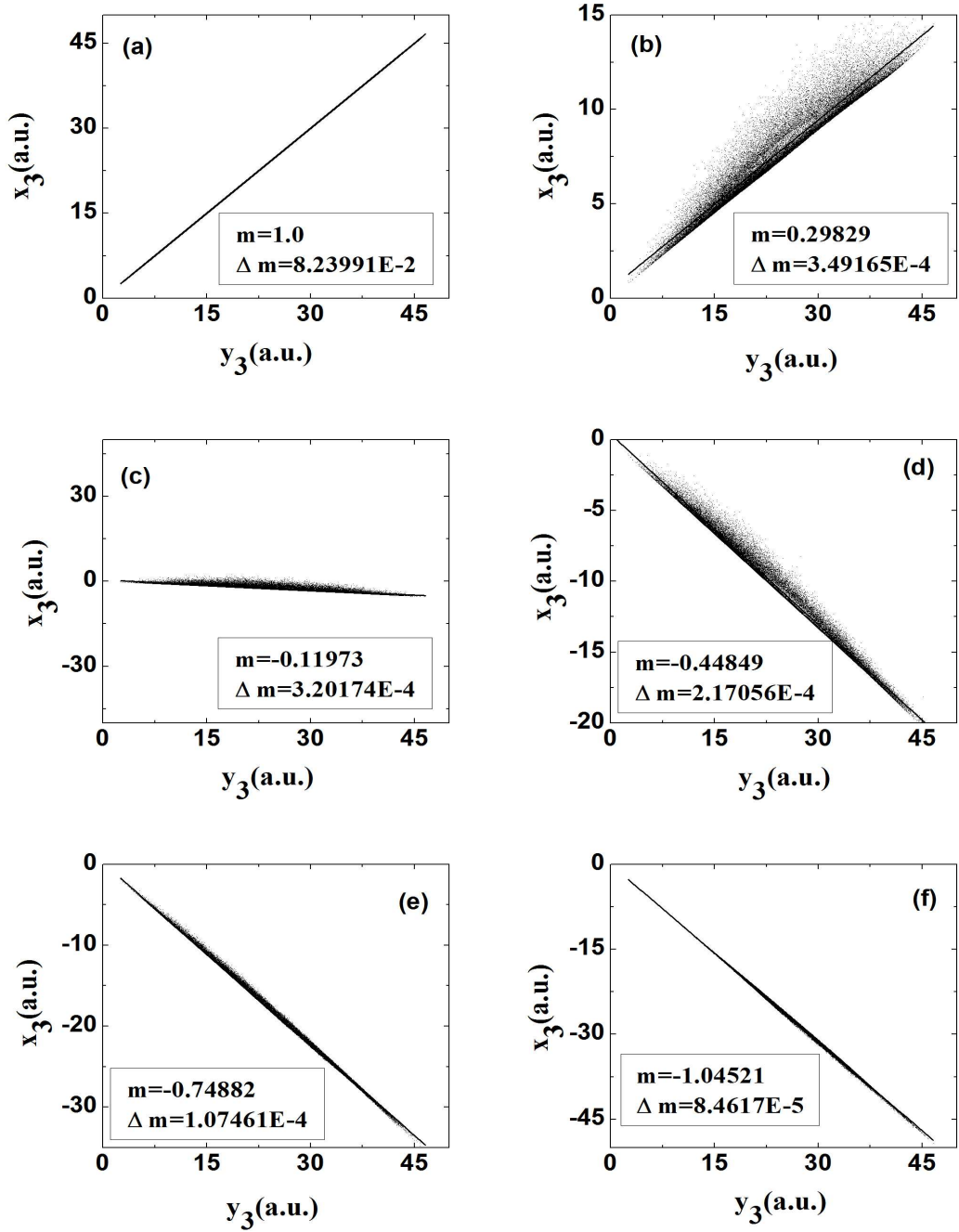

**Fig.3** Synchronization plots plotted between $x_3$ and $y_3$, $\sigma$ is varied from 0.0. to 1.0 (a) $\sigma=0.0$;(b) $\sigma=0.2$;(c) $\sigma=0.4$;(d) $\sigma=0.6$;(e) $\sigma=0.8$;(f) $\sigma=1.0$.

The state of synchronization realized in figures 3(b-e) can be termed as intermediate states of synchronization (IS). The inverse of variance (1/ Δm) gives a measure for quantifying synchronization ($S_Q$ = 1/ Δm). Ideally, for synchronization and anti-synchronization, Δm should tend to zero and thus $S_Q$ value goes to infinity.

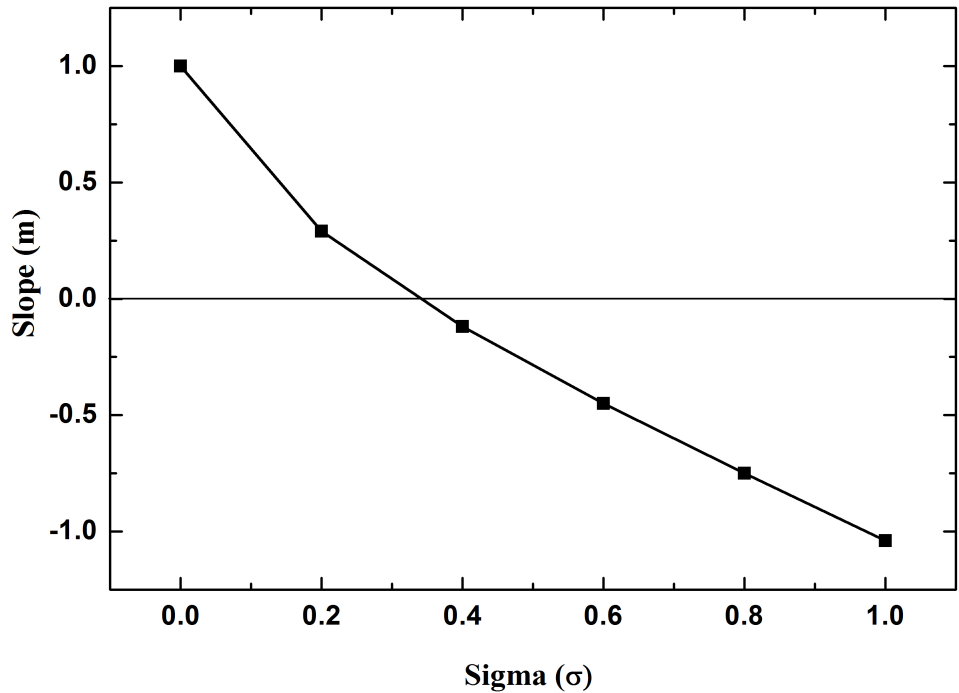

**Fig.4** Slope of linear fit (m) of the synchronization plot as a function of sigma (σ)

The plot of slope (m) vs. $\sigma$ shows a smooth transition from positive synchronization to negative synchronization as shown in figure. 4. Cross - Correlation analysis has been carried out between the $y_3$ and $x_3$ for various values of sigma ($\sigma$) and the results are shown in figure 5. The figure 5(a) and 5(b) displays maxima at zero delay, whereas all the other four plots (figures 5c-f) display minima at zero delay. It is interesting to note that for values of sigma ($\sigma$), resulting in positive slope synchronization plot, the cross-correlation is a maximum at zero delay. But cross-correlation is a minimum for the cases corresponding to negative sloped synchronization plots. Thus the cross-correlation plots affirm the way in which the slave is behaving with respect to the master either positive or negative in its correlation. Hence (i) co-existence of synchronization and anti-synchronization and (ii) the possibility of controlling the sign of synchronization by varying a control parameter of the system are established.

**Application to secure communication**

The GLS system with nonlinear control discussed above is considered for studies of encoding and decoding of a message. Three signals $m_1$, $m_2$ and $m_3$ are transmitted through the three state variables $x_1$, $x_2$ and $x_3$ of the master system respectively. This is achieved by modulating the state variables $x_1$, $x_2$ and $x_3$ using the signals ($m_1$, $m_2$ and $m_3$). This enables encoding of these messages onto the master system. The message signal is defined as, $m_i(t) = a + b \sin f_i(t)$, where; i=1, 2, 3, 'b' is the strength of the signal, 'f' is the frequency of the signal. The signal strength is varied from 0.01 to 0.10. Since the slave is coupled to the master, the messages effectively get transmitted to the slave. Once synchronization (or anti-synchronization) is achieved, the message is decoded by comparing slave's output with that of the master's output. In figure 6, the time traces of master ($x_3$) and slave systems ($y_3$) are shown in (a) and (b) respectively. The decoded signal, which comprises of all of the three signals is shown is figure 6(c) and such presence of all three signals is confirmed from the power spectrum of the decoded signal as shown in figure 6(d). The frequencies of the three signals encoded are 1.000 Hz, 1.088 Hz and 1.250 Hz.

This process is repeated for various combinations of signal frequencies and the results obtained are shown in figures 7, 8, 9 and 10. The different combinations of frequencies considered for encoding are as follows:

Case 1: $f_1 < f_2 < f_3 < f_r$, (figure 7)

Case 2: $f_r < f_1 < f_2 < f_3$, (figure 8)

Case 3: $f_1 < f_2 < f_r$ and $f_3 > f_r$, (figure 9)

Case 4: $f_1 < f_r$ and $f_r > f_2 > f_3$, (figure 10)

where, $f_r$ is the resonance frequency of the GLS system. In all the above four cases, the signal was recovered at the slave by comparing $x_3$ and $y_3$.

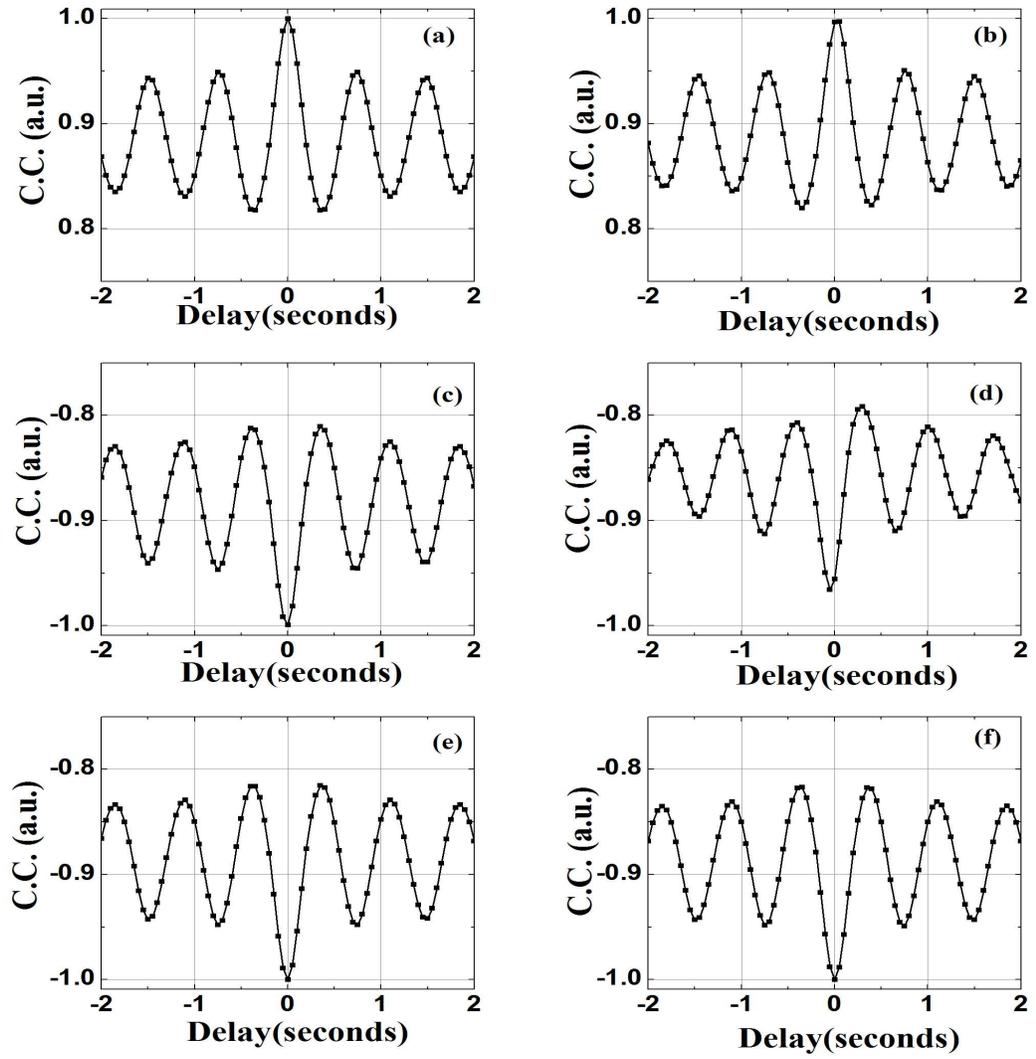

**Fig.5** Cross correlation between $y_3$ and $x_3$ for different values of σ (a) σ=0.0;(b) σ=0.2;(c) σ=0.4;(d) σ =0.6;(e) σ =0.8;(f) σ =1.0.

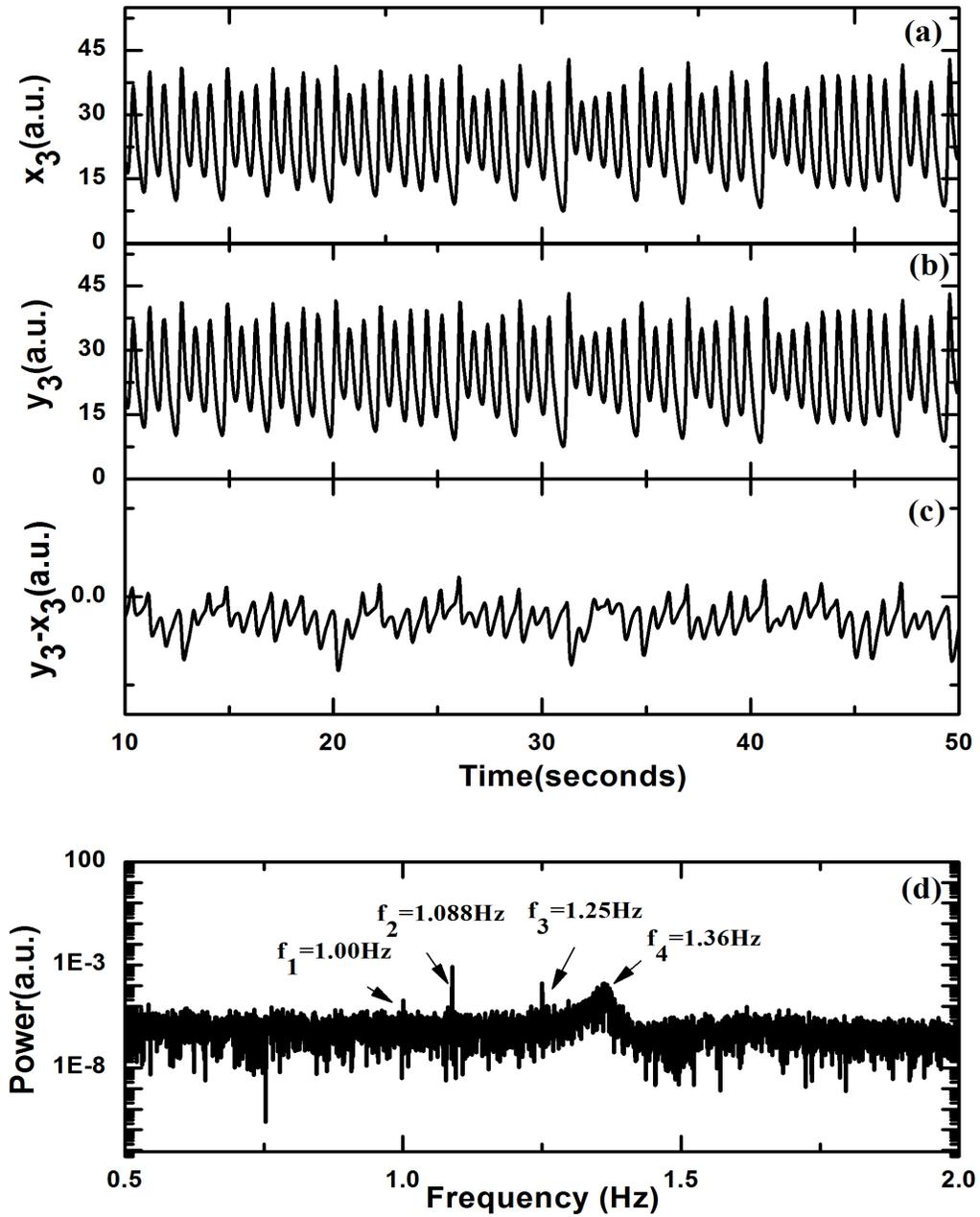

**Fig.6** Message Recovery: (a) Master output; (b) Slave output, (c) error system (recovery);(d) power spectra of error system showing the recovered signals indicated by the arrows. Parameters are: β=0.0 and b=0.003.

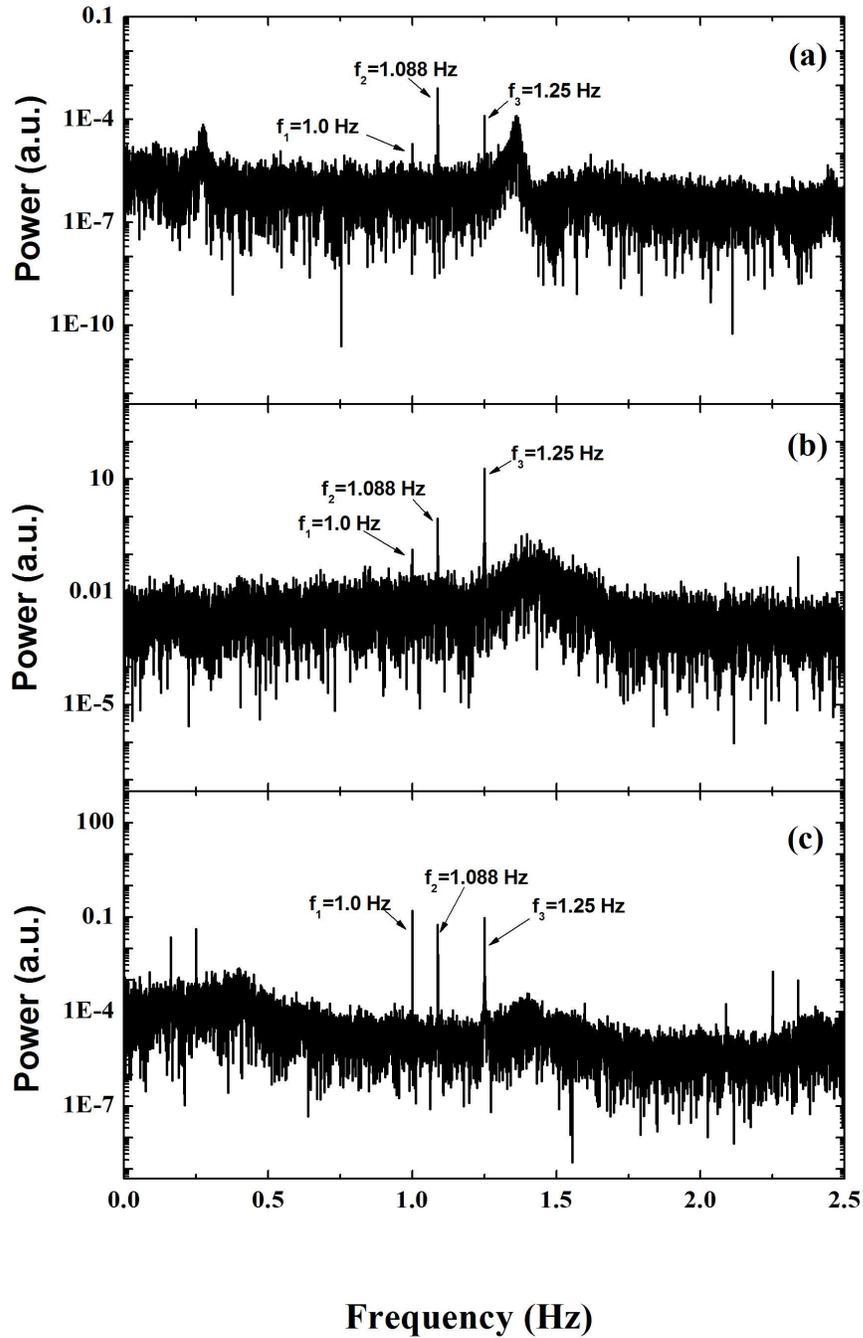

**Fig.7** Power spectra of decoded signal. Case1: $f_1, f_2, f_3 < f_r$; $f_1=1.0$ Hz, $f_2=1.088$ Hz, $f_3=1.25$ Hz, $f_r=1.36$ Hz (a),(b),(c) corresponds to positive slope, zero slope and negative slope respectively to that of the synchronization plot.

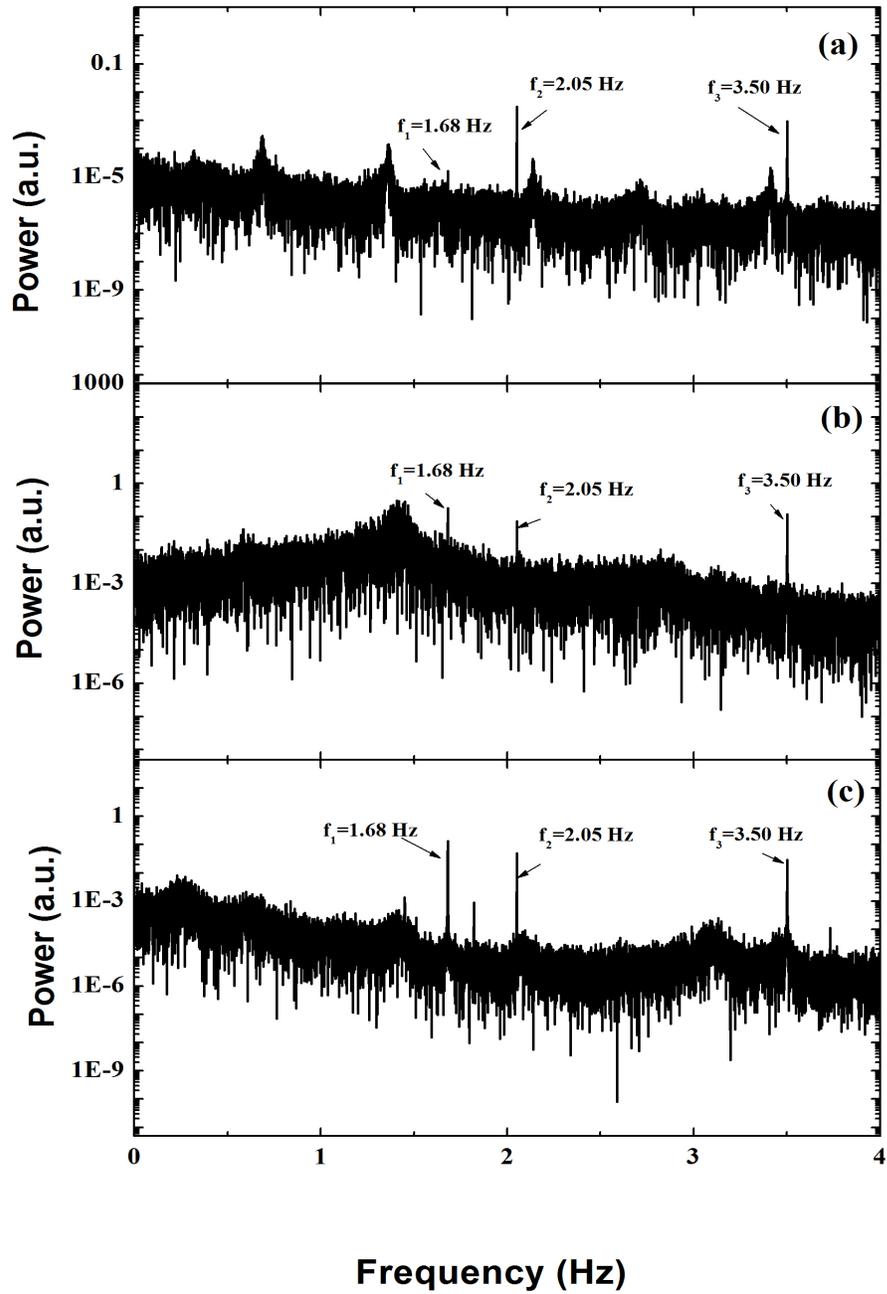

**Fig. 8** Power spectra of decoded signal. Case 2: $f_1, f_2, f_3 < f_r$; $f_1$=1.68 Hz, $f_2$=2.05 Hz, $f_3$=3.50 Hz, $f_r$=1.36 Hz (a),(b),(c) corresponds to positive slope, zero slope and negative slope respectively to that of the synchronization plot.

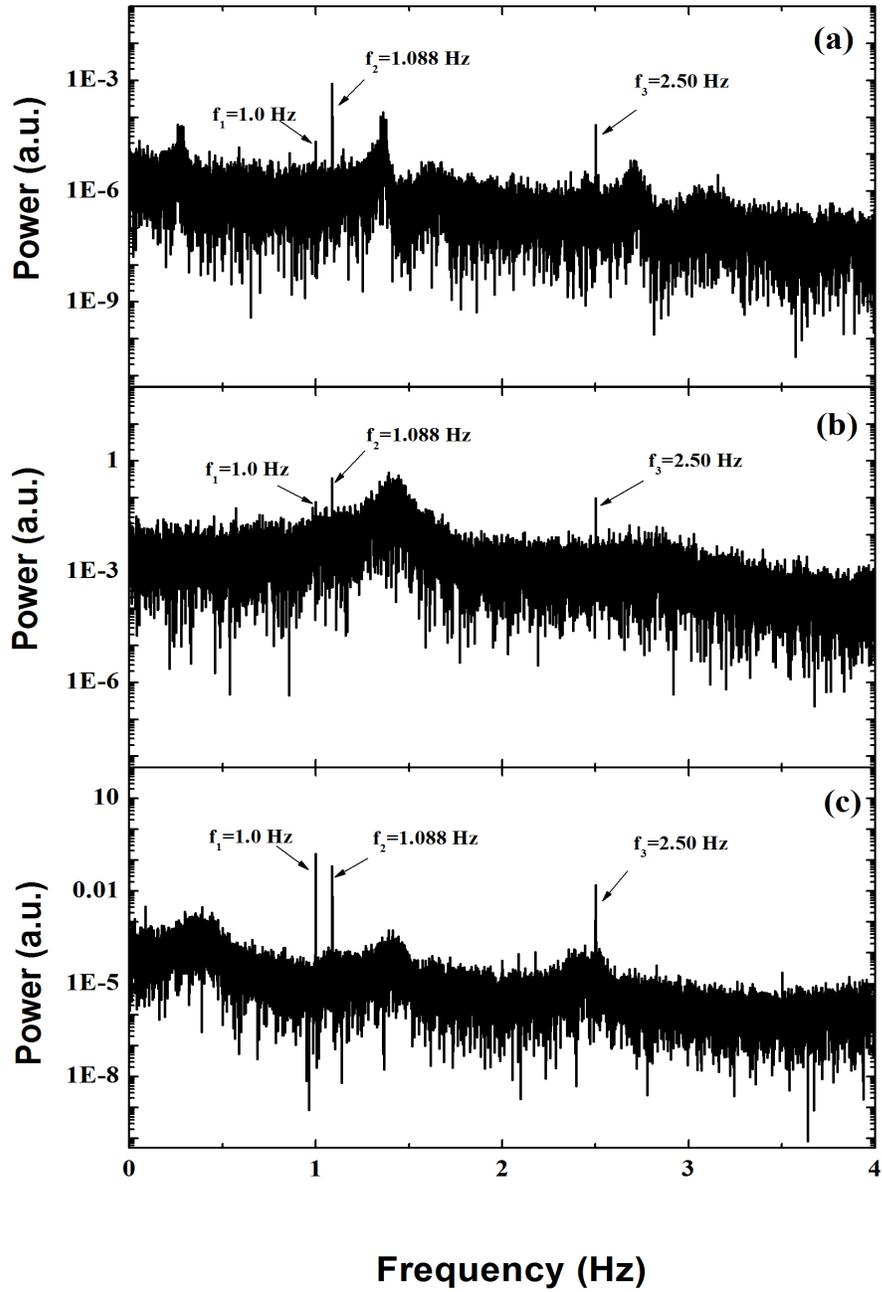

**Fig.9** Power spectra of decoded signal. Case 3: $f_1, f_2 < f_r, f_3 > f_r$; $f_1 = 1.0$ Hz, $f_2 = 1.088$ Hz, $f_3 = 2.50$ Hz, $f_r = 1.36$ Hz (a),(b),(c) corresponds to positive slope, zero slope and negative slope respectively to that of the synchronization plot.

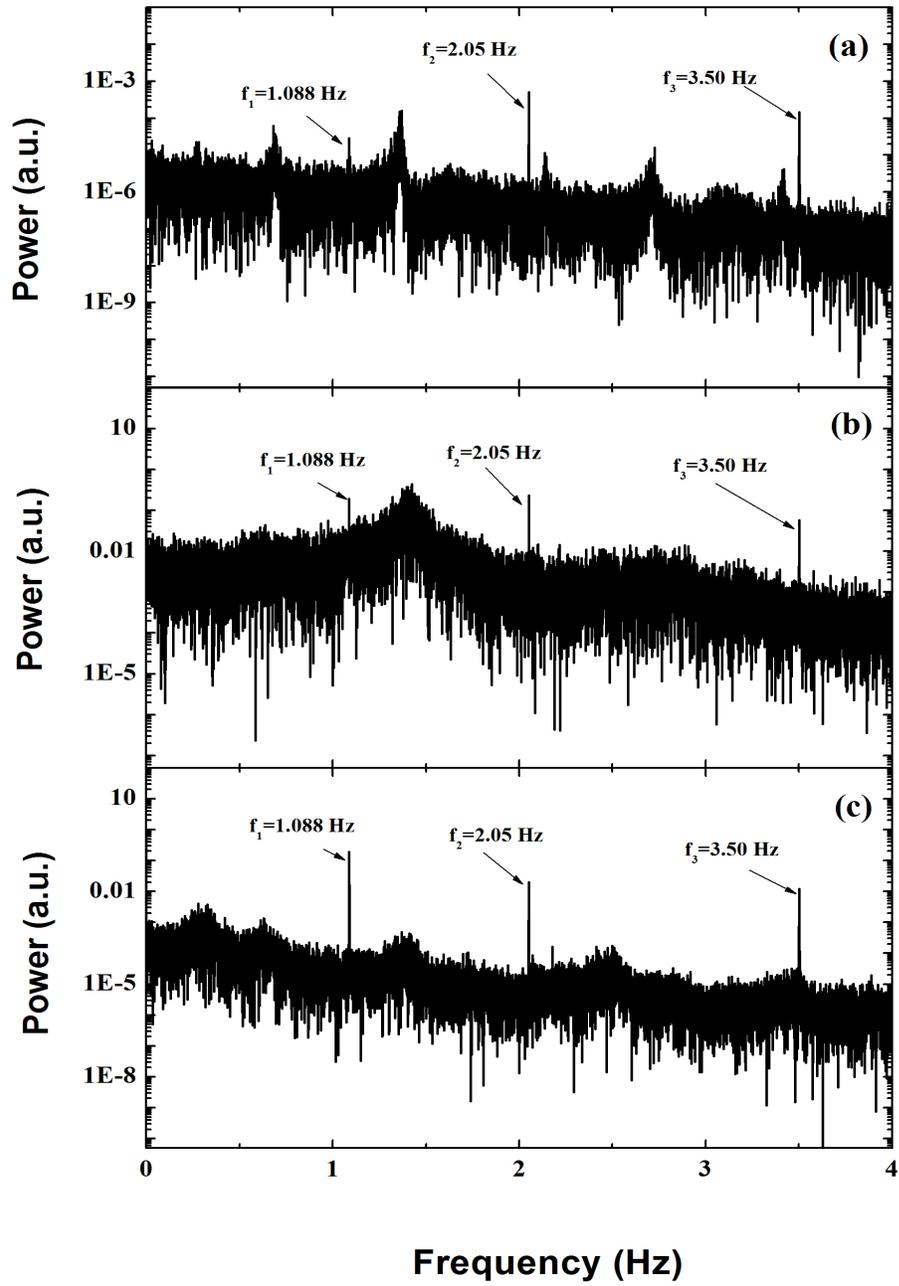

**Fig.10** Power spectra of decoded signal. Case 4: $f_1<f_r, f_2, f_3>f_r; f_1=1.00$ Hz, $f_2=1.088$ Hz, $f_3=2.50$ Hz, $f_r=1.36$ Hz (a),(b),(c) corresponds to positive slope, zero slope and negative slope respectively to that of the synchronization plot.

In figure 7, the power spectra of the decoded signal is shown for the frequencies identified as case-1, for three different values of σ, viz. : σ= 0.0, 0.4 and 1.0, shown as figures 7(a), 7(b) and 7(c) respectively. The sign of synchronization, as detailed in the previous section, changes with the values of sigma. In figure 7 (a, b and c), the presence of encoded signal is obvious and the peaks are identified with arrow marks in the figures.

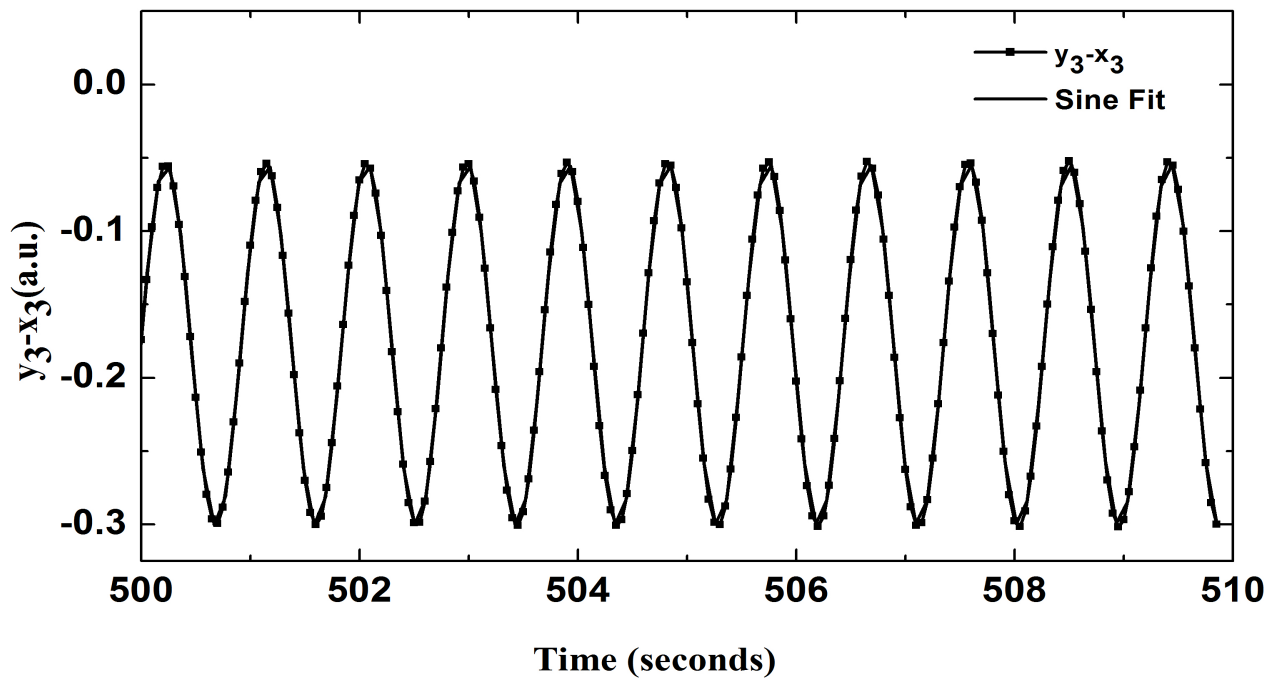

**Fig.11** Sine fit of recovered message. b=0.01; σ=0.0 (synchronization); Adj. R square=0.99997.

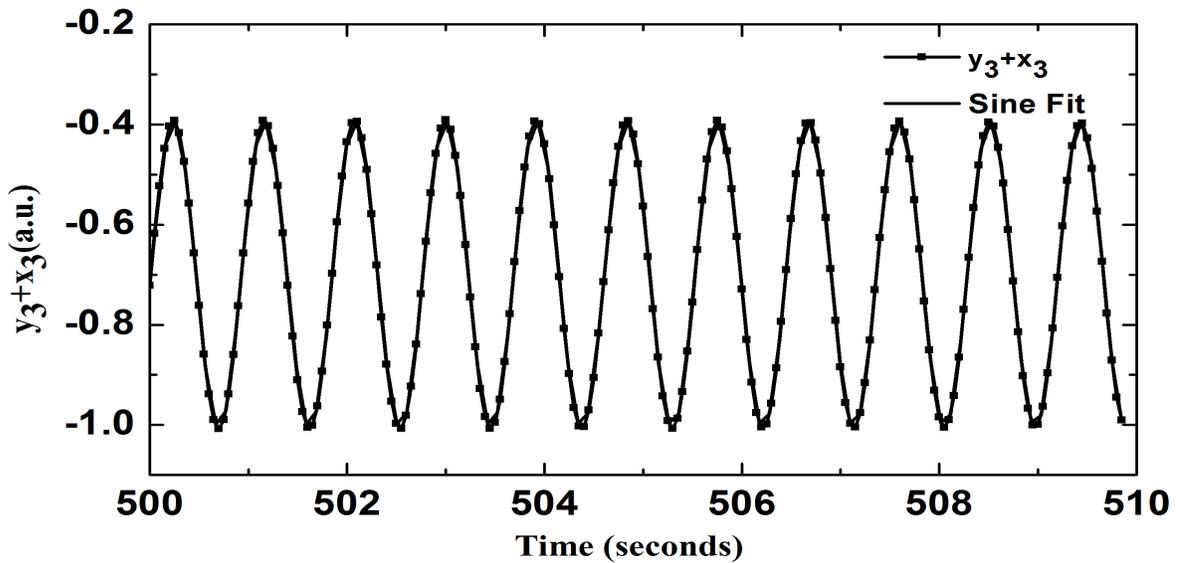

**Fig 12** Sine fit of recovered message. b=0.01; σ=1.0 (anti-synchronization); Adj. R square=0.99998

It is evident that irrespective of the sign of synchronization, decoding is achieved. Figures 8, 9 and 10 show the power spectra of the decoded signal for different frequencies of the encoded signal as considered in case-2, case-3 and case -4 respectively. In all of these figures, three different values of sigma are considered as mentioned in the respective figure captions. In all of these cases, signal decoding is found to be possible and such decoded signals are identified with arrow marks in the figures.

Then the decoded signal is subjected to band pass filtering so as to recover the original encoded signal. For example, in the case-1, the intention is to recover the signal at frequency 1.088 Hz. The time evolving output of error system is subjected to a band pass filter and the resultant filtered output is shown in figure 11 which is evidently a sine character and its frequency is found to be 1.088 Hz. To quantify the quality of signal recovery, the recovered message (obtained by band - pass filtering) is fitted to sine wave and the Adjacent R square value is calculated and is found to be close to unity. This implies high quality recovery of the encoded signal at the slave end. In figure 12, the recovery process for the anti-synchronization regime ($\sigma= 1.0$) is shown and the Adj R square value is found to be close to unity implying a high quality recovery. This process is repeated for all the frequencies and recovery was achieved with the Adj R square value close to unity.

**Conclusions:**

In this work, the synchronization properties of Generalized Lorenz Systems (GLS) arranged in a master-slave configuration is studied. The coupling of master-slave was made possible through a non-linear control mechanism. Control parameters are obtained theoretically such that the error functions are asymptotically stable. Simulations are carried out and synchronization properties between the master-slave's state variables were studied. Anti-synchronization has been achieved between the master and slave for two of the state variables ($x_1$, $y_1$) and ($x_2$, $y_2$). However, the third state variable ($x_3$, $y_3$), upon controlling of parameter exhibits a continuous change-over from synchronization to anti-synchronization via intermediate states of synchronization, while the synchronization between the state variables ($x_1$, $y_1$) and ($x_2$, $y_2$) always remained the in the state of anti-synchronization. Encoding and decoding of messages in this system was carried out. Message is encoded at the master and decoded and recovered at the slave for both states of synchronization, viz. synchronized and anti-synchronized states of master and slave. Simultaneous encoding of messages with distinct frequencies through three state variables of master was carried out and all the three messages were recovered at one of the state variable ($y_3$) of the slave for both the states of synchronization.